\documentstyle[11pt,epsf,curves]{article}
\catcode`\@=11
\def\marginnote#1{}

\newcount\hour
\newcount\minute
\newtoks\amorpm
\hour=\time\divide\hour by60 \minute=\time{\multiply\hour by60
\global\advance\minute by-\hour}
\edef\standardtime{{\ifnum\hour<12 \global\amorpm={am}%
        \else\global\amorpm={pm}\advance\hour by-12 \fi
        \ifnum\hour=0 \hour=12 \fi
        \number\hour:\ifnum\minute<10 0\fi\number\minute\the\amorpm}}
\edef\militarytime{\number\hour:\ifnum\minute<10 0\fi\number\minute}

\def\draftlabel#1{{\@bsphack\if@filesw {\let\thepage\relax
      \xdef\@gtempa{\write\@auxout{\string
          \newlabel{#1}{{\@currentlabel}{\thepage}}}}}\@gtempa \if@nobreak
    \ifvmode\nobreak\fi\fi\fi\@esphack} \gdef\@eqnlabel{#1}}
    \def\@eqnlabel{}
\def\@vacuum{}
\def\draftmarginnote#1{\marginpar{\raggedright\scriptsize\tt#1}}

\def\draft{
  \oddsidemargin -.5truein
  \def\@oddfoot{\footnotesize \sl preliminary draft \hfil
    \rm\thepage\hfil\sl\today\quad\militarytime}
  \let\@evenfoot\@oddfoot \overfullrule 3pt
    \let\label=\draftlabel
    \let\marginnote=\draftmarginnote
  \def\@eqnnum{(\theequation)\rlap{\kern\marginparsep\tt\@eqnlabel}%
    \global\let\@eqnlabel\@vacuum}

  }

\hoffset= - 1.0in

\def\be{\begin{equation}}
\def\ee{\end{equation}}
\def\bea{\begin{eqnarray}}
\def\eea{\end{eqnarray}}
\def\<{\langle}
\def\>{\rangle}

\def\res{{{\rm res}}}

\def\F{{\cal F}}

\def\d{\partial}
\def\N2{${\cal N}=2$}

\def\tr{{\mathrm{tr\,}}}

\def\1N{${\cal N}=1$}
\def\4N{${\cal N}=4$}

\def\e{{\,\rm e}\,}

\def\bea{\begin{eqnarray}}
\def\eea{\end{eqnarray}}
\def\bqa{\begin{eqnarray}}
\def\eqa{\end{eqnarray}}

\def\beq{\begin{equation}}
\def\eeq{\end{equation}}
\def\ba{\beq\begin{array}{c}}
\def\ea{\end{array}\eeq}
\def\be{\beq}
\def\ee{\eeq}

\let\text=\mathrm

\let\wht=\widehat
\let\ov=\overline

\newcommand\theTag[1]{(\ref{#1})}
\def\e{e}

\def\beq{\begin{equation}}
\def\eeq{\end{equation}}
\def\bea{\begin{eqnarray}}
\def\eea{\end{eqnarray}}

\newcommand{\rf}[1]{(\ref{#1})}

\def\F {{\cal F}}

\renewcommand{\d}{{{\partial}}}

\renewcommand{\<}{\langle}
\renewcommand{\>}{\rangle}

\def\2{{1\over 2}}

\def\d{\partial}

\def\â{$\tau$}

\newcommand{\cpict}[3]{
\dimen1=#1\advance\dimen1 by-\hsize\divide\dimen1 by-2 \vtop to #2{
\noindent\hskip\dimen1{\special{em:graph #3.bmp}} \vfil}\hskip-2cm }

\newcommand{\dV}{\frac{\partial}{\partial V(p)}}

\newcommand{\cI}{\oint_{\cal C_{\cal D}}\frac{d\lambda}{2\pi i}}
\newcommand{\Vp}{V^{\prime}}

\newcommand{\ty}{{\tilde y}}

\let\@@savethanks\thanks
\def\thanks#1{\gdef\thefootnote{\alph{footnote}}\@@savethanks{#1}}

\title{{\bf Hermitian matrix model free energy: Feynman graph technique for all
genera } \vspace{.5cm}}
\author{{\bf L. Chekhov}\thanks{E-mail: \ chekhov@mi.ras.ru}
\date{ } \\ {\small
{\it Steklov Mathematical Institute, ITEP}, and {\it LIFR, Moscow, Russia}}\\
and\\
{\bf B. Eynard}\footnote{E-mail: \ eynard@saclay.cea.fr }\ \
\date{ } \\
{\small {\it SPhT, CEA, Saclay, France} }}

\begin{document}
\maketitle

\vspace{-10.5cm}

\begin{center}
\hfill SPhT-T05/046\\
\hfill ITEP/TH-??/05\\
\end{center}

\vspace{7.5cm}

\begin{abstract}
We present the diagrammatic technique for calculating the free
energy of the Hermitian one-matrix model to all orders of $1/N$
expansion in the case where the limiting eigenvalue distribution
spans arbitrary (but fixed) number of disjoint intervals (curves).
\end{abstract}
\def\thefootnote{\arabic{footnote}}

\section{Introduction\label{s:in}}

Matrix models and their so-called multisupport (multicut) solutions
became again important recently due to
studies in ${\cal N}=1$\footnote{We reserve $N$
to denote the size of matrix in matrix models.} \ SUSY gauge
theories by Cachazo, Intrilligator and Vafa \cite{CIV1},
\cite{CIV2} followed by the proposal of Dijkgraaf and Vafa \cite{DV} to
calculate the low energy superpotentials using the partition
function of multicut solutions. These solutions, being
known already for a long time~\cite{JU90,AkAm}, got a new insight
in \cite{David}.

The leading order of the
$1/N$-expansion in the matrix size
of the matrix model is described by the
semiclassical tau-function of the so-called universal Whitham
hierarchy \cite{Kri} (see the
details about one-matrix and two-matrix cases in \cite{ChM} and
\cite{KM}).


One may also consider more general solutions to matrix models, identifying
them with generic solutions to the loop (Schwinger--Dyson, or Virasoro)
equations \cite{Loop}, to be the Ward identities satisfied by matrix integrals
\cite{Virasoro}.

We consider solutions to the loop equations admitting
multi-matrix integral representations
\cite{David,KMT,AMM}. These solutions are associated with families
of Riemann surfaces and form a sort of a basis in the space of all
solutions to the loop equations \cite{AMM} (like the finite-gap
solutions form a similar basis in the space of all solutions to an
integrable hierarchy). They can be distinguished by their
``isomonodromic" properties---switching on higher matrix model
couplings and $1/N$-corrections does not change the family of
Riemann surfaces, but just reparameterizes the moduli as functions
of these couplings.
On the side of integrable hierarchies, these solutions must
satisfy some equations of multicomponent hierarchies \cite{DZ}, not
just the Toda chain hierarchy equations, which are satisfied only by a one-cut
solution.

Much progress has been made on the way to constructing explicit solutions of
matrix models in the large-$N$ expansion since~\cite{BIPZ}. It was shown
(primarily for the one-cut case~\cite{ACKM}) that the variables convenient
for describing solutions in the genus expansion are branching points of the
corresponding Riemann surface and the so-called moments of the potential.
These variables, being expressed highly nonlinearly in times and occupation
numbers, allow reducing the problem to solving some algebraic relations.
Using this technique, the two-matrix model was solved in the large-$N$ limit
in the one-cut case~\cite{Eynard}, this technique was extended to the multicut
case~\cite{KM},\cite{Bertola} , and the subleading term of $1/N$-expansion was found in~\cite{EKK}.

In~\cite{Ey}, the consistent diagrammatic technique for finding the loop means
(multiresolvents) in the $1/N$-expansion of the Hermitian one-matrix model
(1MM) was proposed. The main result of the present paper is the
generalization of this technique to describe the free energy contributions
of all orders in $1/N$ for multicut solutions of the 1MM.

In Sec.~\ref{s:1MM}, we describe the general properties of multi-cut solutions of
1MM and the solution of the loop equation in the large-$N$ limit.

In Sec.~\ref{s:diagram}, we first present necessary notions of the Riemann
geometry describing the Bergmann bidifferentials and the loop insertion operator
(which acts from the space of $s$-differentials to the space of $(s+1)$-differentials)
and then describe the diagrammatic technique of~\cite{Ey} in a more convenient terms
than in the original presentation.

In Sec.~\ref{s:free}, we introduce the operator $H$, which
reduces the degree of forms by one and is in a sense inverse to the loop insertion operator.
This eventually makes possible to present the genus~$k$ contribution to the free energy
through the action of $H$ on the one-point resolvent (for all~$k$ except $k=1$; for the
latter case, the answer has been already obtained, see~\cite{Chekh},~\cite{EKK}).

\section{Matrix models in $1/N$-expansion \label{s:1MM}}

\subsection{Loop equation and resolvents \label{ss:resolvent}}

Let us consider the formal integral of the 1MM~\cite{FGZ}
\be
\int_{N\times N}DX\, \e^{-{N\over t_0}\tr V(X)}=\e^{-\cal F},
\label{X.1}
\ee
where $V(X)=\sum_{n\geq 1}^{}t_nX^n$ and $\hbar = {t_0/ N}$ is
a formal expansion parameter. The integration in (\ref{X.1}) goes
over $N\times N$ Hermitian matrices, and for generic potentials, the
integration may go over curves in the complex plane of each
of $N$ proper variables, which are eigenvalues of $X$.
The topological expansion of the Feynman diagrams series is then equivalent to
the expansion in even powers of $\hbar$ for
\be
{\cal F}\equiv {\cal F}(\hbar,t_0, t_1, t_2, \dots)
=\sum_{g=0}^{\infty}{\hbar}^{2g-2}{\cal F}_g.
\label{X.2}
\ee
Customarily, $t_0=\hbar N$ is the scaled
number of eigenvalues. It is convenient, but not compulsory, to
assume the potential $V(p)$ to be a polynomial
of the fixed degree $m+1$. In general, to have an algebraic curve we need
$V'(p)$ to be a rational function~\cite{Ey1}

The averages corresponding to partition function~\theTag{X.1} are
defined as usual:
\beq
\bigl\langle F(X)\bigr\rangle=
\frac1Z\int_{N\times N}DX\,F(X)\,\exp\left(-{1\over \hbar}\tr V(X)\right),
\label{X.3*}
\eeq
and it is convenient to use their formal
generating functionals: the one-point resolvent
\beq
W(p)=
\hbar
\sum_{k=0}^{\infty}
\frac{\langle\tr X^{k}\rangle}{p^{k+1}}
\label{X.3}
\eeq
as well as the $s$-point resolvents $(s\geq2)$
\beq
W(p_1,\dots,p_s)=
\hbar^{2-s}
\sum_{k_1,\dots,k_s=1}^{\infty}
\frac{\langle\tr X^{k_1}\cdots\tr X^{k_s}\rangle_{\mathrm{conn}}}
{p_1^{k_1+1}\cdots p_s^{k_s+1}}=
\hbar^{2-s}
\left\langle\tr\frac{1}{p_1-X}\cdots
\tr\frac{1}{p_s-X}\right\rangle_{\mathrm{conn}}
\label{X.4}
\eeq
where the subscript ``$\mathrm{conn}$" pertains to the connected
part.

These resolvents are obtained from the free energy ${\cal F}$ through the
action
\bea
W(p_1,\dots,p_s)&=&-\hbar^2\frac{\d}{\d V(p_s)}\frac{\partial}{\partial V(p_{s-1})}\cdots
\frac{\partial {\cal F}}{\partial V(p_1)}=\nonumber
\\
&=&\frac{\partial }{\partial V(p_s)}\frac{\partial }{\partial V(p_{s-1})}\cdots
\frac{\partial }{\partial V(p_2)}W(p_1),
\label{X.5}
\eea
of the loop insertion operator\footnote{Although this operator contains all
partial derivatives w.r.t. the variables
$t_k$'s, below we introduce additional variables
$S_i$ and, therefore, we use the
partial derivative notation here.}
\beq
\frac{\partial }{\partial V(p)}\equiv
-\sum_{j=1}^{\infty}\frac{1}{p^{j+1}}\frac{\d}{\d t_{j}}.
\label{X.6}
\eeq
Therefore, if one knows exactly the one-point resolvent for arbitrary
potential, all multi-point resolvents can be calculated by induction.
In the above normalization, the genus expansion has the form
\beq
W(p_1,\dots,p_s)=\sum_{g=0}^{\infty}
\hbar^{2g}
W_{g}(p_1,\dots,p_s),\quad s\geq1,
\label{X.7}
\eeq
which is analogous to genus expansion \rf{X.2}.

The first in the chain of the loop equations of the 1MM is~\cite{Loop}
\beq
\oint_{{\cal C}_{\cal D}}\frac{d\omega}{2\pi i}\frac{V'(\omega)}{p-\omega}W(\omega)=
W(p)^2+
\hbar^2
W(p,p).
\label{X.8}
\eeq
Here and hereafter, ${\cal C}_{\cal D}$~is a contour encircling clockwise
all singular points
of $W(\omega)$, but not the point
$\omega=p$; this contour integration acts as the projection operator
extracting the coefficient
of the term~$p^{-1}$. Using Eq.~\theTag{X.5}, one can express
the second term in the r.h.s.\ of loop equation~\theTag{X.8} through
$W(p)$, and Eq.~\theTag{X.8} becomes an equation on
one-point resolvent \rf{X.3}.

Substituting genus expansion~\theTag{X.7} in Eq.~\theTag{X.8}, we find
that $W_g(p)$ for $g\geq1$ satisfy the equation
\beq
\left(\widehat{K}-2W_{0}(p)\right)W_g(p)=\sum_{g'=1}^{g-1}
W_{g'}(p)W_{g-g'}(p)+\frac{\partial }{\partial V(p)}W_{g-1}(p),
\label{X.9}
\eeq
where $\widehat{K}$~is the linear integral operator
\beq
\widehat{K}f(p)\equiv-\oint_{{\cal C}_{\cal D}}\frac{d\xi}{2\pi i}
\frac{V'(\xi)}{p-\xi}f(\xi).
\label{X.10}
\eeq
In Eq.~\theTag{X.9}, $W_g(p)$ is expressed through only the
$W_{g_i}(p)$ for which $g_i<g$. This fact permits
developing the iterative procedure.

The solution $W_1(p)$ to the loop equation in the multicut case
was first found by
Akemann~\cite{Ak96}.\footnote{The universal critical behavior of the
corresponding correlation functions was discussed in~\cite{AkAm}.}
He also managed to integrate it in order to obtain the free energy $\F_1$ in the
two-cut case.
The genus-one partition function in the generic multi-cut
case was proposed in \cite{Kos,DST}, where it was observed that the Akemann formula
coincides with the correlator of twist fields (that produce cuts on the
complex plane and give rise to the hyperelliptic
Riemann surface as a cover of complex plane) computed by Al.Zamolodchikov
\cite{Zam}, with some corrections due to including star operators
(introduced by G.Moore in \cite{Moore}) into consideration.
In~\cite{Chekh,EKK}, the genus-one correction was derived
by solving the loop equation, which
generalize the Akemann result for partition function to arbitrary number
of cuts.

\subsection{Solution in genus zero} \label{ss:basic}

The form of loop equation (\ref{X.8}) is based exclusively on the
reparameterization invariance of the matrix integral, which retains
independently on the details of eigenvalue density distribution.
In the 1MM case at $N\to\infty$, the eigenvalues fill in some segments
in complex plane, dependently on the shape of potential $V(X)$. For
polynomial potentials, the number of segments is finite and
the contour ${\cal C_{\cal D}}$ of integration in (\ref{X.9})
encircles a finite number~$n$ of disjoint intervals
\be
{\cal D} \equiv \bigcup_{i=1}^n [\mu_{2i-1},\mu_{2i}],
\quad \mu_1< \mu_2< \ldots < \mu_{2n}.
\ee
Recall that as
\be
W_g(p)|_{p\to\infty} = \frac{t_0}{p}\delta_{g,0}+O({1}/{p^2}),
\label{Winf}
\ee
all $W_g(p)$ must be total derivatives,
\be
\label{total}
W_g(p)=\dV {\cal F}_g,\quad g\ge 0.
\ee
Inserting genus expansion (\ref{X.7}) into loop equation
(\ref{X.8}), we obtain
\be
\cI \frac{V^{\prime}(\lambda)}{p-\lambda} W_0(\lambda) = (W_0(p))^2
\label{plan}
\ee
for genus zero and (\ref{X.9}) for higher genera.
Given $W_0(p)$, one can then determine $W_g(p)$ for $g\geq 1$ iteratively
genus by genus provided the operator $(\widehat{K}-2W_0(p))$ (see (\ref{X.10}))
can be inverted uniquely.

One can solve Eq. (\ref{plan}) for the planar one-point resolvent $W_0(p)$
as follows. Deforming the contour in Eq.~(\ref{plan}) to infinity, we obtain
\be
(W_0(p))^2  = \Vp(p) W_0(p)
 +\oint_{\cal C_{\infty}}\frac{d\lambda}{2\pi i} \frac{\Vp(\lambda)}{p-\lambda}W_0(\lambda).
 \label{*loop1*}
\ee
The last term in the r.h.s. is
\be
\label{polP}
P_{m-1}(p)  = \oint_{\cal C_{\infty}}\frac{d\lambda}{2\pi i}
      \frac{\Vp(\lambda)}{p-\lambda}W_0(\lambda),
\ee
a degree-$m-1$ polynomial to be fixed later, and the solution to \rf{*loop1*}
is then
\be
W_0(p) = \frac{1}{2}\Vp(p) - \frac{1}{2}\sqrt{\Vp(p)^2+4P_{m-1}(p)}
\equiv \frac{1}{2}\Vp(p) - y(p),
 \label{*loop2*}
\ee
where the minus sign is chosen in order to fulfill asymptotic condition (\ref{Winf})
and the function $y(p)$ is defined as follows.
For the polynomial potential of power $m+1$, the resolvent $W_0(p)$ is
a function on complex plane with $m$ cuts, or on a hyperelliptic curve
$y^2 = \Vp(p)^2+4P_{m-1}(p)$ of genus $g=m-1$.
For generic potential $V(X)$ with $m\to\infty$, this curve
may have an infinite genus, but
we can still consider solutions with a finite, fixed number $n$ of cuts
filled by eigenvalues. For this, we separate the smooth part of
the curve introducing
\be
\label{ty}
y(p)\equiv M(p)\ty(p), \quad \hbox{and} \quad
{\ty}^2(p)\equiv\prod\nolimits_{\alpha=1}^{2n}(p-\mu_\alpha)
\ee
with all branching points $\mu_\alpha$ distinct. The variable $\ty$ defines therefore
the new, reduced Riemann surface, which plays a fundamental role in our construction.
In what follows, we still assume
$M(p)$ to be a polynomial of degree $m-n$, keeping in mind that $n$ is always
finite and fixed, while $m\geq n$ can be chosen arbitrarily large.

From now on, we distinguish between images of the infinity at two sheets---physical
and unphysical---of hyperelliptic Riemann surface (\ref{ty}) respectively denoting them
$\infty_+$ and $\infty_-$.
By convention, we set $\ty|_{p\to\infty_+}\sim p^{n}$, and
$M(p)$ is then\footnote{By a standard convention, $\res_{\infty}dx/x=-1$, and the
direction of the integration contour at the infinity point therefore coincides with the
direction of contour for integrals over $\cal C_{\cal D}$ and over the set of $A$-cycles, see below.}
\be
M(p) =-\frac12 \res_{\infty_+} {dw} \frac{\Vp(w)}{(w-p)\ty(w)}.
\label{M}
\ee
Inserting this solution in Eq. (\ref{*loop2*}) and deforming
the contour back, we obtain the planar one-point resolvent
with an $n$-cut structure,
\be
W_0(p) = \frac{1}{2}\cI \frac{\Vp(\lambda)}{p-\lambda}\frac{\ty(p)}{\ty(\lambda)},
\quad p\not\in{\cal D}.
\label{W0}
\ee

Let us now discuss the parameter counting. Rewriting \rf{*loop1*} as quadratic
equation for $W_0(p)$ we had to imply that polynomial \rf{polP} depends on some
unknown parameters to be fixed later. Indeed, the dependence on coefficients of \rf{polP}
can be retranslated into the dependence on {\em filling fractions}
\be
\label{Sfr}
S_i =
\oint_{A_i}\frac{d\lambda}{2\pi i}\,y(\lambda)=\oint_{A_i}\frac{d\lambda}{2\pi i}M(\lambda)\ty(\lambda),
\ee
where $A_i$, $i=1,\dots,n-1$ is the basis of $A$-cycles on the hyperelliptic
Riemann surface \rf{ty} (we may conveniently choose them to be the first $n-1$ cuts).
Adding the (normalized) total number of eigenvalues
\be
\label{t0}
t_0 = \oint_{\cal C_{\cal D}}y(\lambda) \frac {d\lambda}{2\pi i} = \res_{\infty_+}\, y(\lambda)d\lambda
\ee
to the set of $S_i$, we obtain $n$ parameters, which exactly matches
the number of coefficients of \rf{polP}
in the nondegenerate case where $m=n$.
We assume the
occupation number for the last, $n$th cut to be $t_0-\sum_{i=1}^{n-1}S_i\equiv S_n$.
\footnote{It is sometimes convenient to consider $S_n$ instead of $t_0$ as a canonical variable.
However, in all instants we use $S_n$, we specially indicate it for not confusing $S_n$ with the ``genuine''
filling fraction variables $S_i$, $i=1,\dots,n-1$.}
If $n<m$, we consider variables \rf{Sfr} as conditions
to which we add, following (\ref{Winf}), the asymptotic conditions
\be
-t_0\delta_{k,n}  =  \frac{1}{2}
 \cI \frac{\lambda^k V^{\prime}(\lambda)}{\ty(\lambda)}, \quad k=0,\ldots,n,
 \label{Rand1}
\ee
which provide $n+1$ equations for $2n$ constants $\mu_\alpha$.

In the planar limit of matrix models, filling fractions \rf{Sfr}
can be considered as {\it independent\/} parameters of the theory, since the
jumps between different cuts are suppressed as non-perturbative corrections in $\hbar$.
(Obviously, no parameters $S_i$ arise in the one-cut case.)
In particular, this imposes restrictions
\be
\label{dVSi}
\dV S_i=0,\quad i=1,\dots,n-1,\qquad \dV t_0=0.
\ee
In accordance with the original
matrix model formulation (see e.g. \cite{JU90},~\cite{Ak96}) one must look
for the genuine
minimum of the matrix model effective action with respect to {\em all} variables:
$S_i$ together with the times $t_j$. This implies vanishing of partial derivatives of
${\cal F}_0$ with respect to $S_i$, and one can find (see~\cite{JU90},~\cite{DV},
and~\cite{ch-rev}) that
these derivatives are differences of chemical potential on disjoint cuts,
equal to the integrals over {\em dual\/} $B$-cycles on \rf{ty}:
\be
\label{Dvdual}
{\d {\cal F}_0\over\d S_i} = \oint_{B_i}y(\lambda)d\lambda \equiv \Pi_i
\ee
Note that the geometric definition of ${\cal F}_0$ is {\em not} modular
invariant, i.e., it depends on the choice of $A$- and $B$-cycle
basis on \rf{ty}.
Under the change of homology basis, ${\cal F}_0$ transforms in accordance with the
duality transformations \cite{dWM}.
The higher-genus corrections become also scheme dependent: choosing
$S_i$ or $\Pi_i$ as independent variables, we obtain
different expressions for the genus-one free energy.

\section{Calculating resolvents. Diagram technique  \label{s:diagram}}

In this section, we rederive the diagrammatic technique of~\cite{Ey} in a more
concise form of differentials on Riemann surface (\ref{ty}). We choose $S_i$
to be independent variables.
Our main goal
is to invert loop equation (\ref{X.9}) to obtain the expression for $W_k(p)$
for any $k\ge1$.

\subsection{Bergmann bidifferential and two-point resolvent \label{ss:B}}

Let us consider the
{\it Bergmann kernel} (canonically normalized bidifferential in Fay's terminology)
which is the bi-differential on a Riemann
surface~$\Sigma_g$ that is symmetrical in its arguments $P,Q\in \Sigma_g$
and has the only singularity at the coinciding arguments where it has the
behavior (see~\cite{Fay},~\cite{Kra})
\be
B(P,Q)=\left(\frac{1}{(\tau(P)-\tau(Q))^2}+\frac16S_B(P)+o(1)\right)d\tau(P)d\tau(Q),
\label{*Bergmann*}
\ee
in some local coordinate $\tau(P)$; $S_B(P)$ is the Bergmann projective connection
and we fix the normalization (the possibility to add symmetrical bilinear
forms composed from holomorphic 1-differentials in variables $Q$ and $P$) claiming
vanishing all the integrals over $A$-cycles of $B(P,Q)$:
\be
\oint_{A_i}B(P,Q)=0,\ \hbox{for}\ i=1,\dots,g.
\label{*vanish*}
\ee
We then have the following standard Rauch variational formulas relating $B(P,Q)$ with other objects on
a (general, not necessarily hyperelliptic) Riemann surface:
\be
\frac{\d}{\d\mu_\alpha}B(P,Q)=\frac12 B(P,[\mu_\alpha])B([\mu_\alpha],Q),
\label{dB}
\ee
and
\be
\oint_{B_i}B(P,Q)=2\pi i dw_i(P),
\label{B-int}
\ee
where $\mu_\alpha$ is any simple branching point of the complex structure, then, by definition,
in the vicinity of $\mu_\alpha$,
\be
B(P,Q)|_{Q\to\mu_\alpha}=B(P,[\mu_\alpha])\left(\frac{dq}{\sqrt{q-\mu_\alpha}}+O(\sqrt{q-\mu_\alpha})dq\right),
\label{Y.1}
\ee
and $dw_i(P)$ are canonically normalized holomorphic differentials:
\be
\oint_{A_j}dw_i(P)=\delta_{ij}.
\label{dw}
\ee

We also introduce the 1-form $dE_{Q,Q_0}(P)$, which is the primitive of $B(P,Q)$:
\be
dE_{Q,Q_0}(P)=\int_{Q_0}^QB(P,\xi),\qquad dE_{Q,Q_0}(P)|_{P\to Q}=\frac{d\tau(P)}{\tau(P)-\tau(Q)}+\hbox{finite\,}.
\label{dE}
\ee
Then, obviously,
\be
\oint_{A_i}dE_{Q,Q_0}(P)=0.
\label{dEA}
\ee
The form $dE_{Q,Q_0}(P)$ is single-valued w.r.t. $P$ on the Riemann surface and multiple-valued w.r.t. the
variable $Q$: from (\ref{B-int}),
\bea
&&dE(P)_{Q+\oint_{B_i},Q_0}(P)=2\pi i dw_i(P)+dE_{Q,Q_0}(P),
\nonumber
\\
&&dE(P)_{Q+\oint_{A_i},Q_0}(P)=dE_{Q,Q_0}(P).
\nonumber
\eea
The reference point $Q_0$ does not play any role in the construction; we keep it only
for the consistency of notation.

These quantities are related to the Prime form $E(P,Q)$:
\be
B(P,Q)=d_Pd_Q\log E(P,Q),\qquad dE_{Q,Q_0}(P)=d_P\log\frac{E(P,Q)}{E(P,Q_0)},
\ee
where the Prime form is defined in the standard way.
Let us consider the Jacobian~$J$, which is a $g$-dimensional torus
related to the curve~$\Sigma_g$. Recall that the Abel map
$\Sigma_g\mapsto J:\ Q\to \vec x\equiv\left\{\int_{Q_0}^Qd\omega_i\right\}$, where $Q_0$ is a
reference point, set into the correspondence to each point $Q$ of the complex curve
the vector in the Jacobian, and we also introduce the theta function
$\Theta_{[\alpha]}(\vec x)$ of an odd characteristic $[\alpha]$
that becomes zero at $\vec x=0$. We introduce the normalizing functions
(1/2-differentials) $h_{\alpha}(x)$ determined for the subset of $\vec x\in J$ that are
image points of $Q\in\Sigma_g$
$$
h^2_{\alpha}(\vec x)=\sum_{i=1}^g\frac{\d \Theta_{[\alpha]}}{\d z_i}(0)d\omega_i(Q).
$$
The explicit expression for the
Prime form $E(x,y)$ that has a single zero on the Riemann surface~$\Sigma_g$
then reads
\be
E(P,Q)=\frac{\Theta_{[\alpha]}(\vec x-\vec y)}{h_{\alpha}(\vec x)h_{\alpha}(\vec y)}.
\label{*E(P,Q)*}
\ee

We can now express the 2-point resolvent
$W_0(p,q)$ in terms of $B(P,Q)$. Let us denote $p$ and $\ov p$ the points on the
respective physical and unphysical sheets. Then,
$$
\frac{\d V'(p)}{\d V(q)}=-B(p,q)-B(p,\ov q)=-\frac{dp\,dq}{(p-q)^2}
$$
since it has double poles with unit quadratic residues at $p=q$ and $p=\ov q$.
The 2-point resolvent (\ref{*loop2*}) is nonsingular at coinciding points; therefore,
\be
\frac{\d y(p)}{\d V(q)}=-\frac12 (B(p,q)-B(p,\ov q)),
\label{dVy}
\ee
and
\be
W_0(p,q)=-B(p,\ov q).
\label{W2}
\ee

\subsection{The iterative procedure}\label{ss:iterate}

We can determine
higher genus contributions iteratively by inverting
genus expanded loop equation (\ref{X.9}). All
multi-point resolvents of the same genus can be obtained from $W_g(p)$ merely
by applying the loop insertion operator $\dV$.

Claiming all the higher free energy terms $F_g$ to depend only on $\mu_\alpha$
and a {\em finite} number of the moments $M_\alpha^{(k)}$, which are derivatives
of $(k-1)$th order of the polynomial $M(p)$ at branching points, allows
no freedom of adding the terms depending only on $t_0$ and $S_i$ to $F_g$.

We first prove that the operator $\wht K-2W_0(p)$ acting on
\be
\oint_{{\cal C}_{\cal D}}dE_{q,q_0}(p)\frac{dq}{2\pi i}\frac{1}{2y(q)}f(q),
\label{Y.2}
\ee
where $f(q)$ is the combination of resolvents in the r.h.s. of (\ref{X.9})
whose only singularities are poles of finite orders at $\mu_\alpha$ and
the point $p$ lies outside ${\cal C}_{\cal D}$, just gives $f(p)$.

For this, we first note that as $V'(\xi)$ has no square root singularities at $\mu_\alpha$,
 instead of evaluating the integral over ${\cal C}_{\cal D}$,
we can just evaluate {\em residues} on Riemann surface (\ref{ty}) w.r.t. the variables
$\xi$ and $q$. (This effectively corresponds to doubling the integral over ${\cal C}_{\cal D}$
and it gives additional factor $1/2$.)
We therefore denote ${\cal C}^{(\xi)}$ the contour that encircles all branching
points of $\ty(\xi)$ and assume that the contour ${\cal C}^{(q)}$
of integration w.r.t. the variable $q$ lies inside ${\cal C}^{(\xi)}$.

We now implement the Riemann bilinear identities. The contributions of $A$- and $B$-cycle
integrations vanish due to (\ref{dVSi}) and (\ref{B-int}).
Following (\ref{*loop2*}), we substitute $2(W_0(\xi)+y(\xi))$ for $V'(\xi)$ in
(\ref{X.10}) and calculate the part with $W_0(\xi)$ taking the contour of integration
to the infinity; the integrand with $dE_{q,q_0}(\xi)$ is regular, and the only contribution
comes from poles at $\xi=p$ and $\xi=\ov p$;
this contribution exactly cancels the term $-2W_0(p)$.

The remaining integral is
$$
-\oint_{p>{\cal C}^{(\xi)}>{{\cal C}^{(q)}}}\frac{d\xi}{2\pi i}\frac{dq}{2\pi i}
\frac{y(\xi)}{p-\xi}\frac{\ty(p)}{\ty(\xi)}dE_{q,q_0}(\xi)\frac{f(q)}{2y(q)}.
$$
Since $y(\xi)/\ty(\xi)=M(\xi)$ is regular at $\xi=\mu_\alpha$, pushing the
integration contour ${\cal C}^{(\xi)}$ through ${{\cal C}^{(q)}}$ we obtain zero upon
the integration over $\xi$, and the only nonzero contribution comes from the simple pole
of $dE_{q,q_0}(\xi)\sim 1/(\xi-q)$ at $\xi=q$. Terms with $y(q)$ in the numerator and
denominator then cancel each other and the remaining integral w.r.t. $q$ can be taken pushing
the contour again to infinity evaluating the residues at $q=p$ and $q=\ov p$, which gives
$\frac12(f(p)+f(\ov p))=f(p)$ for a function analytic in $p$ outside the branching points.

This provides a basis for the diagrammatic representation for resolvents in 1MM~\cite{Ey}.
Let us represent the form $dE_{q,q_0}(p)$ as the vector directed
from $p$ to $q$, the three-point vertex as the dot in which
we assume the integration over $q$, \ $\bullet\equiv\oint\frac{dq}{2\pi i}\frac{1}{2y(q)}$, and the
Bergmann 2-form $B(p,q)$ as a nonarrowed edge connecting points $p$ and $q$. The graphic representation
for a solution of (\ref{X.9}) then looks as follows.
Representing the multiresolvent $W_{g'}(p_1,\dots,p_s)$ as the block with $s$ external legs and with the index $g'$,
we obtain~\cite{Ey}
\be
\begin{picture}(190,55)(10,10)
\thicklines
\put(40,40){\oval(20,20)}
\thinlines
\put(20,40){\line(1,0){12}}
\put(20,40){\circle*{2}}
\put(20,44){\makebox(0,0)[cb]{$p$}}
\put(40,40){\makebox(0,0)[cc]{$g$}}
\put(55,40){\makebox(0,0)[lc]{$=\sum\limits_{g'=1}^{g-1}$}}
\put(80,40){\vector(1,0){13}}
\put(80,40){\circle*{2}}
\put(95,40){\circle*{4}}
\put(80,44){\makebox(0,0)[cb]{$p$}}
\put(93,44){\makebox(0,0)[cb]{$q$}}
\put(95,40){\line(1,1){7}}
\put(95,40){\line(1,-1){7}}
\thicklines
\put(107,52){\oval(20,20)}
\put(107,28){\oval(20,20)}
\put(107,52){\makebox(0,0)[cc]{$g-g'$}}
\put(107,28){\makebox(0,0)[cc]{$g'$}}
\thinlines
\put(120,40){\makebox(0,0)[lc]{$+$}}
\put(130,40){\vector(1,0){13}}
\put(130,40){\circle*{2}}
\put(145,40){\circle*{4}}
\put(130,44){\makebox(0,0)[cb]{$p$}}
\put(143,44){\makebox(0,0)[cb]{$q$}}
\put(145,40){\line(1,1){9}}
\put(145,40){\line(1,-1){9}}
\thicklines
\put(166,40){\oval(30,30)}
\put(166,40){\makebox(0,0)[cc]{$g-1$}}
\put(185,40){\makebox(0,0)[cc]{$,$}}
\end{picture}
\label{Bertrand}
\ee
which provides the diagrammatic representation for $W_k(p_1,\dots,p_s)$. The multiresolvent
$W_k(p_1,\dots,p_s)$ can be presented as a finite sum of all possible connected
graphs with $k$ loops and $s$ external legs and with only three-valent internal vertices
(the total number of edges is then $2s+3k-3$, and we assume $s\ge1$ for $k\ge1$ and $s\ge3$ for $k=0$).
We also set arrows on some (exactly $2k+s-2$) of edges for the arrowed edges to constitute the
rooted tree subgraph with all arrows directed from the root.
That means that we choose one of the external legs, say, $p_1$
(the choice is arbitrary due to the symmetry of $W_k(p_1,\dots,p_s)$), to be
the root vertex the tree starts with; for each three-valent
vertex there must exist exactly one incoming edge of the tree subgraph, all external edges (except the
root edge) are nonarrowed, and all internal nonarrowed edges either start and terminate at the same
vertex (we then associate with such an edge the Bergmann differential $B(p,\ov p)$) or connect
two different vertices such that there exists the directed path composed by arrowed edges connecting
these two vertices. At each internal vertex we have the integration
$\oint_{{\cal C}^{(q)}}\frac{dq}{2\pi i}\frac{1}{2y(q)}$, while the arrangement of the integration contours
at different vertices is prescribed by the arrowed subtree: the closer is a vertex to the root, the more outer
is the integration contour.

We now demonstrate the consistency of this diagram technique by calculating the action of loop insertion
operator (\ref{X.6}) on its elements.

We first calculate the action of $\d/\d V(r)$ on $B(P,Q)$. Using (\ref{dB}), we represent this action through
the action of partial derivatives in $\mu_\alpha$ subsequently calculating the latter from relation (\ref{dVy}).
Let
$$
y(x)dx|_{x\to\mu_\alpha}=y([\mu_\alpha])\sqrt{x-\mu_\alpha}dx+O(\sqrt{x-\mu_\alpha})^3dx.
$$
Then, since
$$
\left.\frac{\d y(p)dp}{\d V(r)}\right|_{p\to\mu_\alpha}\simeq
-\frac12 y([\mu_\alpha])\frac{dp}{\sqrt{p-\mu_\alpha}}\frac{\d\mu_\alpha}{\d V(r)},
$$
we have
\be
\frac{\d \mu_\alpha}{\d V(r)}=\frac{2B([\mu_\alpha],q)}{y([\mu_\alpha]),}
\label{dmul}
\ee
and, therefore
\be
\frac{\d}{\d V(r)}B(P,Q)=\sum_{\alpha=1}^{2n} \frac{B(P,[\mu_\alpha])B([\mu_\alpha],Q)B([\mu_\alpha],r)}{y([\mu_\alpha])}.
\label{Y.4}
\ee
For expressing this in terms of the differentials, we integrate one of the Bergmann bidifferentials in order
to obtain the 1-differential $dE_{Q,Q_0}(\xi)$.\footnote{This differs from the commonly accepted viewpoint that one
should instead take the primitive of $y(\xi)$. But the 1-differential $ydx$, being the Whitham differential $dS$ in
other terminology~\cite{Kri},~\cite{ChM}, plays the fundamental role, not its primitive.} Note that the local
variable in the vicinity of $\mu_\alpha$ is $\xi(x)=\sqrt{x-\mu_\alpha}$, this gives the additional factor $1/2$, and we
eventually obtain
\be
\frac{\d}{\d V(r)}B(P,Q)=\sum_{\alpha=1}^{2n} \res_{\mu_\alpha}\frac{B(P,\xi(x))dE_{\xi(x),Q_0}(Q)B(\xi(x),r)}{2y(xi(x))d\xi(x)}.
\label{Y.5}
\ee
From this relation it obviously follows that
\be
\frac{\d}{\d V(r)}dE_{P,Q_0}(Q)=\sum_{\alpha=1}^{2n} \res_{\mu_\alpha}\frac{dE_{P,Q_0}(\xi(x))dE_{\xi(x),Q_0}(Q)B(\xi(x),r)}{2y(xi(x))d\xi(x)},
\label{Y.6}
\ee
and the last quantity to evaluate is
\be
\frac{\d}{\d V(r)}\frac{1}{2y(p)}=-\frac{B(p,r)}{2y^2(p)}.
\label{Y.7}
\ee

Note that the point $P$ in (\ref{Y.6}) is outside the integration contour. (This is irrelevant in formula (\ref{Y.5}) as the Bergmann
bidifferential has zero residue at the coinciding arguments, in contrast to $dE_{P,Q_0}(Q)$. Multiplying by $1/(2y(P))$ both sides
of (\ref{Y.6}), using (\ref{Y.7}), and pushing the integration contour through the point $P$, we observe that the contribution of the
simple pole of $dE_{P,Q_0}(\xi(x))$ at $\xi(x)=P$ cancels exactly the variation of $1/(2y(P))$. We therefore attain the prescribed contour
ordering and can graphically present the action of $\d/\d V(r)$ as
\be
\begin{picture}(240,55)(10,10)
\thicklines
\put(5,40){\makebox(0,0)[cc]{$\frac{\d}{\d V(r)}$}}
\put(20,40){\makebox(0,0)[cc]{$Q$}}
\put(25,40){\vector(1,0){15}}
\put(45,40){\makebox(0,0)[cc]{$P$}}
\put(55,40){\makebox(0,0)[cc]{$=$}}
\put(65,40){\makebox(0,0)[cc]{$Q$}}
\put(70,40){\vector(1,0){10}}
\put(80,40){\circle*{2}}
\put(80,40){\vector(1,0){10}}
\put(95,40){\makebox(0,0)[cc]{$P,$}}
\put(80,40){\line(0,1){10}}
\put(80,55){\makebox(0,0)[cc]{$r$}}
\put(125,40){\makebox(0,0)[cc]{$\frac{\d}{\d V(r)}$}}
\put(140,40){\makebox(0,0)[cc]{$Q$}}
\put(145,40){\line(1,0){15}}
\put(165,40){\makebox(0,0)[cc]{$P$}}
\put(175,40){\makebox(0,0)[cc]{$=$}}
\put(185,40){\makebox(0,0)[cc]{$Q$}}
\put(190,40){\vector(1,0){10}}
\put(200,40){\circle*{2}}
\put(200,40){\line(1,0){10}}
\put(215,40){\makebox(0,0)[cc]{$P$}}
\put(200,40){\line(0,1){10}}
\put(200,55){\makebox(0,0)[cc]{$r$}}
\put(225,40){\makebox(0,0)[cc]{$\equiv$}}
\put(235,40){\makebox(0,0)[cc]{$Q$}}
\put(240,40){\line(1,0){10}}
\put(250,40){\circle*{2}}
\put(260,40){\vector(-1,0){10}}
\put(265,40){\makebox(0,0)[cc]{$P\ .$}}
\put(250,40){\line(0,1){10}}
\put(250,55){\makebox(0,0)[cc]{$r$}}
\end{picture}
\label{variation}
\ee
In the second case, it is our choice on which of edges to set the arrow. Recall, however, that the points $P$ and $Q$ were already
ordered as prescribed by the diagram technique. That is, if ``$P\to Q$'', we must choose the first variant and if ``$Q\to P$'',
we must choose the second variant of arrows arrangement.

\section{Inverting the loop insertion operator. Free energy  \label{s:free}}

\subsection{The $H$-operator \label{ss:H}}

We now introduce the operator that is in a sense inverse to loop insertion operator (\ref{X.6}).
Let\footnote{This definition works well when acting on 1-forms regular at infinities. Otherwise (say, in the case of
$W_0(p)$), the integral in the third term must be regularized, e.g., by replacing it by the contour integral around the logarithmic
cut stretched between two infinities.}
\be
H\cdot=\frac12\res_{\infty_+}V(x)\cdot\ -\frac12\res_{\infty_-}V(x)\cdot\
-t_0\int_{\infty_-}^{\infty_+}\cdot\ -\sum_{i=1}^{n-1}S_i\oint_{B_i}\cdot.
\label{H}
\ee
The arrangement of the integration contours see in Fig.~1. We now calculate the action of $H$ on the Bergmann
bidifferential $B(x,q)$ using again the Riemann bilinear identities.
We first note that $B(x,q)=\d_xdE_{x,Q_0}(q)$ and we can evaluate residues at infinities
by parts. Then, since $dE_{x,Q_0}(q)$ is regular at infinities, we substitute for $V'(x)$ $2y(x)+2t_0/x$ as $x\to\infty_+$
and $-2y(x)+2t_0/x$ as $x\to\infty_-$ thus obtaining
\bea
&&-\res_{\infty_+}\left(y(x)+\frac{t_0}{x}\right)dE_{x,Q_0}(q)dx+\res_{\infty_-}\left(-y(x)+\frac{t_0}{x}\right)dE_{x,Q_0}(q)dx
\nonumber
\\
&&\qquad \Bigl.-t_0dE_{x,Q_0}(q)\Bigr|_{x=\infty_-}^{x=\infty_+}-\sum_{i=1}^{n-1}S_i\oint_{B_i}B(q,x),
\label{Z.1}
\eea
whence the cancelation of terms containing $t_0$ is obvious, and it remains to take the combination of residues
at infinities involving $y(x)$. For this, we cut the surface along $A$- and $B$-cycles taking into account the residue at
$x=q$. The boundary integrals on two sides of the cut at $B_i$ then differ by $dE_{x,Q_0}(q)-dE_{x+\oint_{A_i},Q_0}(q)=0$,
while the integrals on two sides of the cut at $A_i$ differ by  $dE_{x,Q_0}(q)-dE_{x+\oint_{B_i},Q_0}(q)=\oint_{B_i}B(q,x)$,
and we obtain for the boundary term the expression
$$
\sum_{i=1}^{n-1}\oint_{A_i}y(x)dx\oint_{B_i}B(q,\xi),
$$
which exactly cancel the last term in (\ref{Z.1}). It remains only the contribution from the pole at $x=q$, which is just
$-y(q)$. We have therefore proved that
\be
H\cdot B(\cdot,q)=-y(q)dq.
\label{HB}
\ee

\begin{figure}[tb]
\vskip .2in
\setlength{\unitlength}{0.8mm}%
\begin{picture}(0,40)(15,15)
\thinlines
\put(135,43){\circle{10}}
\put(135,43){\circle*{1}}
\put(133,45){\makebox(0,0)[cc]{$q$}}
\put(134,37.7){\vector(-1,0){0}}
\thicklines
\curve(60,30, 62,33.5, 65,35, 70,36, 80,36.5, 90,36, 95,35, 98,33.5, 100,30, 98,26.5, 95,25, 90,24, 80,23.5, 70,24, 65,25, 62,26.5, 60,30)
\put(80,36.5){\vector(1,0){0}}
\curve(64,29, 66,31, 68,29, 70,31, 72,29, 74,31, 76,29, 78,31, 80,29, 82,31, 84,29, 86,31, 88,29, 90,31, 92,29, 94,31, 96,29)
\put(80,40){\makebox(0,0)[rb]{$A_i$}}
\curve(124,29, 126,31, 128,29, 130,31, 132,29, 134,31, 136,29, 138,31, 140,29, 142,31, 144,29, 146,31, 148,29, 150,31, 152,29, 154,31, 156,29)
\put(110,40){\makebox(0,0)[rb]{$B_i$}}
\curve(92,26.5, 90,30, 92,33.5, 95,35, 100,36, 110,36.5, 120,36, 125,35, 128,33.5, 130,30, 128,26.5)
\put(110,36.5){\vector(1,0){0}}
{\curvedashes[1mm]{0,1,2}
\curve(92,33.5, 90,30, 92,26.5, 95,25, 100,24, 110,23.5, 120,24, 125,25, 128,26.5, 130,30, 128,33.5)}
\put(204,35){\circle*{1}}
\put(204,25){\circle*{1}}
{\thinlines
\curvedashes[1mm]{1,3,2}
\curve(204,35, 157,35, 154,34, 153,33, 152,30, 153,27, 154,26, 157,25, 204,25)
\curvedashes[1mm]{5,1}
\curve(204,35, 157,35, 154,34, 153,33, 152,30, 153,27, 154,26, 157,25, 204,25)}
\thinlines
\put(208,35){\makebox(0,0)[cc]{${\scriptstyle \infty_+}$}}
\put(208,25){\makebox(0,0)[cc]{${\scriptstyle \infty_-}$}}
\put(208,32){\vector(1,0){0}}
\put(208,22){\vector(1,0){0}}
\curve(153.5,30, 154,31.8, 154.5,32.8, 155.5,33.5, 157.5,34, 164,34, 199,34, 200,34, 201,33.5, 205,32, 209,32, 210,32.3, 211,32.8, 212,34, 212.3,35)
\curve(150.5,30, 151,32.2, 151.5,33.7, 152.5,35, 154.5,36, 162,36, 199,36, 200,36, 201,36.5, 205,38, 209,38, 210,37.7, 211,37.2, 212,36, 212.3,35)
\curvedashes[0.5mm]{2,1}
\curve(153.5,30, 154,28.2, 154.5,27.2, 155.5,26.5, 157.5,26, 164,26, 199,26, 200,26, 201,26.5, 205,28, 209,28, 210,27.7, 211,27.2, 212,26, 212.3,25)
\curve(150.5,30, 151,27.8, 151.5,26.3, 152.5,25, 154.5,24, 162,24, 199,24, 200,24, 201,23.5, 205,22, 209,22, 210,22.3, 211,22.8, 212,24, 212.3,25)
\end{picture}
\caption{The arrangement of integration contours on the Riemann surface.}
\label{fi:cuts}
\end{figure}
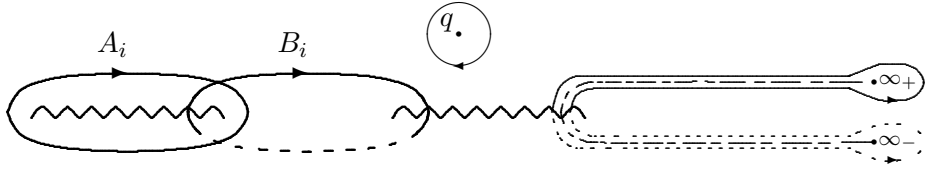

Let us now consider the action of $H$ on $W_k(\cdot)$ subsequently evaluating the action of loop insertion operator (\ref{X.6}) on the
result. Note first that the only result of action of $\d/\d V(p)$ on the operator $H$ itself are derivatives
$\d V(x)/\d V(p)=-1/(p-x)$ (and recall that by definition $|p|>|x|$, i.e., instead of evaluating residues at infinities one should
take residues at $x=p$, and we obtain
\be
\frac{\d}{\d V(p)}\left(H\cdot W_k(\cdot)\right)=W_k(p)+H\cdot W_k(\cdot,p).
\label{Z.2}
\ee
For the second term, due to the symmetry of $W_k(p,q)$, we may take the point $p$ as the root of the tree subgraph. Then,
the operator $H$ always acts on $B(\cdot,\xi)$ where $\xi$ are integration variables of internal vertices. However, if this vertex
is an innermost (i.e., there is no arrowed edges coming out of it), then the 1-form $y(\xi)d\xi$ arising under the action of $H$
(\ref{HB}) cancels the corresponding form in the integration expression, and the residue vanishes being nonsingular at the
branching point. If there is an outgoing arrowed edge, say $dE_{\rho,Q_0}(\xi)$
(can be just one as one line is external), then, again, we can push the integration contour for $\xi$ through the one for $\rho$;
the only contribution comes again only from the pole at $\xi=\rho$. The value of the residue is doubled, and we come to the
following graphical representation for the action of the operator $H$:
\be
\begin{picture}(240,30)(10,30)
\thicklines
\put(25,40){\makebox(0,0)[cc]{$Q$}}
\put(30,40){\vector(1,0){10}}
\put(40,40){\circle*{2}}
\put(40,40){\vector(1,0){10}}
\put(55,40){\makebox(0,0)[cc]{$P$}}
\put(40,40){\line(0,1){10}}
\put(40,55){\makebox(0,0)[cc]{$H\cdot$}}
\put(65,40){\makebox(0,0)[cc]{$=$}}
\put(75,40){\makebox(0,0)[cc]{$-$}}
\put(85,40){\makebox(0,0)[cc]{$Q$}}
\put(90,40){\vector(1,0){15}}
\put(110,40){\makebox(0,0)[cc]{$P$}}
\put(116,40){\makebox(0,0)[cc]{;}}
\put(135,40){\makebox(0,0)[cc]{$Q$}}
\put(140,40){\vector(1,0){10}}
\put(150,40){\circle*{2}}
\put(150,40){\line(1,0){10}}
\put(165,40){\makebox(0,0)[cc]{$P$}}
\put(150,40){\line(0,1){10}}
\put(150,55){\makebox(0,0)[cc]{$H\cdot$}}
\put(175,40){\makebox(0,0)[cc]{$=$}}
\put(185,40){\makebox(0,0)[cc]{$0.$}}
\end{picture}
\label{chopping}
\ee
For $H_q\cdot W_k(q,p)=H_q\cdot \frac{\d}{\d V(q)}W_k(p)$, we obtain that for each arrowed edge on which the action of
(\ref{X.6}) produces the new vertex, the inverse action of $H_q\cdot$ just give the factor $-1$ and on each nonarrowed edge
on which the action of (\ref{X.6}) produces the new vertex, the inverse action of $H_q\cdot$ just gives zero. As the total number
of arrowed edges is $2k-1$ for any graph contributing to the sum of diagrams, we obtain that
$$
H_q\cdot W_k(q,p)=-(2k-1)W_k(p)
$$
and, combining with (\ref{Z.2}), we just obtain
\be
\frac{\d}{\d V(p)}\left(H_q\cdot W_k(q)\right)=-(2-2k)\frac{\d}{\d V(p)}F_k,
\label{Z.3}
\ee
and, since all the dependence on filling fractions and $t_0$ is fixed by the claim that the answer depends only on $\mu_\alpha$ and
derivatives of $M(p)$ of finite orders at the branching points, we conclude that
\be
F_k=\frac{1}{2k-2}H\cdot W_k.
\label{fin}
\ee
This is our final answer for the free energy. It permits us to calculate all $F_k$ except the contribution at $k=1$ (torus approximation).
It was however calculated by a direct integration in~\cite{Chekh}. All other orders can be consistently calculated. For this,
we only introduce one new vertex $\circ$ in which we have $\oint_{{\cal C}^{(\xi)}}\frac{d\xi}{2\pi i}\frac{\int_{\mu_\alpha}^\xi y(s)ds}{y(\xi)}$.
Although the integral term $\int_{Q_0}^\xi y(s)ds$ is nonlocal, its constant part $\int_{Q_0}^{\mu_\alpha}$ drops out of the residue in the
1MM case, and we can integrate it in the neighborhoods of every branching point $\mu_\alpha$ separately. That is, we need only the expansion
of this integral as $\xi\to\mu_\alpha$. Then, say, the genus two contribution is provided by the sum of three diagrams
\be
\begin{picture}(240,30)(10,30)
\thicklines
\put(15,40){\makebox(0,0)[cc]{$2F_2$}}
\put(30,40){\makebox(0,0)[cc]{$=$}}
\put(38,40){\makebox(0,0)[cc]{$2$}}
\put(45,40){\circle{3}}
\put(50,40){\circle{10}}
\put(55,40){\circle*{2}}
\put(53,45){\vector(1,0){0}}
\put(55,40){\vector(1,0){10}}
\put(65,40){\circle*{2}}
\put(70,40){\circle{10}}
\put(85,40){\makebox(0,0)[cc]{$+2$}}
\put(95,40){\circle{3}}
\put(105,40){\circle{20}}
\put(105,50){\circle*{2}}
\put(105,50){\vector(0,-1){20}}
\put(98.5,47.5){\vector(1,1){0}}
\put(105,30){\circle*{2}}
\put(125,40){\makebox(0,0)[cc]{$+$}}
\put(135,40){\circle{10}}
\put(140,40){\circle*{2}}
\put(150,40){\vector(-1,0){10}}
\put(150,40){\circle{3}}
\put(150,40){\vector(1,0){10}}
\put(160,40){\circle*{2}}
\put(165,40){\circle{10}}
\end{picture}
\label{F2}
\ee

For completeness, we also present
$$
F_0=\frac{1}{2\pi i}\oint_{{\cal C}_{\cal D}}V(p)y(p)dp-\frac{1}{(2\pi i)^2}\oint_{{\cal C}_{\cal D}}\oint_{{\cal C}_{\cal D}}y(p)y(q)\log|p-q|dp\,dq
$$
which can also be written like in (\ref{fin})
\be
F_0=-\frac{1}{2}H_{\rm reg}\cdot W_0.
\ee
where $H_{\rm reg}$ is explained in many articles (see \cite{Bertola}),
and
\be
F_1=-\frac{1}{24}\log\left(\prod_{\alpha=1}^{2n}M(\mu_\alpha)\Delta^4(\det\sigma)^{12}\right),
\label{F1}
\ee
where $\Delta$ is the Vandermonde determinant for $\mu_\alpha$ and $\sigma$ is $(n-1)\times(n-1)$-matrix
$$
\sigma_{ij}=\oint_{A_i}\frac{\xi^{j-1}d\xi}{\ty(\xi)};
$$
note that $\det\sigma=\det^{-1}{\widehat H}_{ij}$, where ${\widehat H}_{ij}$ are coefficients of the polynomials expressing
canonically normalized holomorphic 1-differentials:
$$
dw_i=\frac{{\widehat H}_i(\xi)}{\ty(\xi)}d\xi,
$$
and, for instance,
$$
dE_{q,q_0}(\xi)-dE_{\bar q,q_0}(\xi)=\frac{\ty(q)d\xi}{(\xi-q)\ty(\xi)}-\sum_{i=1}^{n-1}\frac{{\widehat H}_i(\xi)d\xi}{\ty(\xi)}
\oint_{A_i}\frac{d\rho\ty(q)}{(\rho-q)\ty(\rho)}.
$$
Meanwhile, it is easy to see that $H\cdot W_1$ is constant. Indeed,
$$
H\cdot W_1=\sum_{\alpha=1}^{2n}\res_{\mu_\alpha}\frac{\int_{\mu_\alpha}^\xi y(s)}{y(\xi)}B(\xi,\bar\xi)d\xi,
$$
the first term has simple zero with residue $1/2$ to be compensated by the double pole of $B(\xi,\bar\xi)\simeq \frac{1}{4(\xi-\mu_\alpha)^2}$,
and the total answer is then just the constant $n/4$.

\subsection{Scaling relation \label{ss:scaling}}

We now a little demystify relation (\ref{Z.3}). Indeed, recall that for any functional $F$ of a finite
number of ``local'' variables, which are in our case the branching points $\mu_\alpha$ and
the moments
$$
\left.M_\alpha^{(k)}=\frac{1}{(k-1)!}\frac{d^{k-1}}{dp^{k-1}}M(p)\right|_{p=\mu_\alpha},
$$
we have the relations
\be
\frac{\d F}{\d t_0}=\int_{\infty_-}^{\infty_+}\frac{\d F}{\d V(\xi)}d\xi
\label{Z.4}
\ee
and
\be
\frac{\d F}{\d S_i}=\oint_{B_i}\frac{\d F}{\d V(\xi)}d\xi.
\label{Z.5}
\ee
Relation (\ref{Z.3}) is then equivalent to the relation
\be
\left[\sum_{k=1}^\infty t_k\frac{\d}{\d t_k}+t_0\frac{\d}{\d t_0}+\sum_{i=1}^{n-1}S_i\frac{\d}{\d S_i}+\hbar\frac{\d}{\d \hbar}\right]F=0
\label{base}
\ee
satisfied by the total 1MM free energy. Indeed, coming back to basics of the matrix model, note that the
operator in (\ref{base}) just generates the scaling transformations: the part with derivatives w.r.t. $t_k$ multiplies all vertices
by the same scaling factor $\rho$ simultaneously multiplying propagators by $\rho^{-1}$. The action of derivatives in $t_0$ and $S_i$
results in multiplying all index loops (faces of the fat graph) by $\rho$. Therefore, for any graph, the total factor is
$$
\rho^{\hbox{\scriptsize \#\ vertices -- \#\ edges + \#\ faces}}=\rho^{2-2g},
$$
and it is exactly canceled by the scaling of the formal expansion parameter $\hbar\to\rho\hbar$.

\section*{Acknowledgments}

Our work is partly supported by the RFBR grant No.~05-01-00498 (L.Ch.),
by the Grant of Support for the Scientific
Schools NSh-2052.2003.1 (L.Ch.),
and by the Program Mathematical Methods of Nonlinear Dynamics (L.Ch.).
B.E. thanks ENIGMA European Network MRTN-CT-2004-5652.

\end{document}